\documentclass{aa}  

\usepackage[colorlinks,allcolors=blue]{hyperref}

\usepackage{graphicx}
\usepackage{subcaption}
\usepackage{txfonts}

\makeatletter
\renewcommand*\aa@pageof{, page \thepage{} of \pageref*{LastPage}}
\makeatother

\defcitealias{GilFab20}{GF20}

\begin{document}

   \title{Quantifying the Similarity of Planetary System Architectures}

   \subtitle{}

   \author{D.~Bashi
          \inst{1}
          \and
          S.~Zucker\inst{1}}

   \institute{Porter School of the Environment and Earth Sciences, Raymond and Beverly Sackler Faculty of Exact Sciences, Tel Aviv University, Tel Aviv, 6997801, Israel.\\
              \email{dolevbas@mail.tau.ac.il ; shayz@tauex.tau.ac.il}
             }

   \date{Received ... ; accepted ...}

\abstract {The planetary systems detected so far already exhibit a wide diversity of architectures, and various methods are proposed to study quantitatively this diversity. Straightforward ways to quantify the difference between two systems and more generally, two sets of multiplanetary systems, are useful tools in the study of this diversity. In this work we present a novel approach, using a Weighted extension of the Energy Distance (WED) metric, to quantify the difference between planetary systems on the logarithmic period-radius plane. We demonstrate the use of this metric and its relation to previously introduced descriptive measures to characterise the arrangements of \textit{Kepler} planetary systems. By applying exploratory machine learning tools, we attempt to find whether there is some order that can be ascribed to the set of \textit{Kepler} multiplanet system architectures. Based on WED, the 'Sequencer', which is such an automatic tool, identifies a progression from small and compact planetary systems to systems with distant giant planets. It is reassuring to see that a WED-based tool indeed identifies this progression. Next, we extend WED to define the Inter-Catalogue Energy Distance (ICED) -- a distance metric between sets of multiplanetary systems. We have made the specific implementation presented in the paper available to the community through a public repository. We suggest to use these
metrics as complementary tools in attempting to compare between architectures of planetary system, and in general, catalogues of planetary systems.
}


   \maketitle
%

\section{Introduction} \label{sec:Intro}

One of the most prominent results by the \textit{Kepler} mission \citep{Boruckietal10} was the detection of hundreds of multiplanet systems.
The analysis of the architectures of these systems revealed the ubiquity of close-in sub-Neptune, super-Earth and Earth-like planets, and led to important insights about the statistical relations among planets within such systems \citep{Lissaueretal11, Fabryckyetal14, Weissetal18}. One example is the "peas in a pod" pattern, first noted by \citet{Milletal17,Weissetal18}, and further discussed in \citet{Kipping18, WeissPet20, Muldersetal20, Heetal20}. In systems that exhibit this pattern, planets orbiting the same host tend to be similar in size and have regular orbital spacing. At present, it is not completely clear yet whether this pattern is indeed of astrophysical origin or a selection effect due to observational biases \citep{Zhu20, MurchikovaTremaine20},
nonetheless, the evidence seems to support the case that the pattern is of astrophysical origin \citep{WeissPet20, GilFab20, Heetal20}.

There are currently two main approaches that can be found in the literature to analyze quantitatively the architectures of planetary systems:

The first approach uses various attributes of the system architecture that are easy to identify and quantify. Some authors used information theory in order to identify such attributes. Thus, \cite{Kipping18} proposed a model to define the entropy of the planet size ordering within a planetary system. Using this model, he then argued that the observed \textit{Kepler} multiplanet systems displayed a highly significant deficit in entropy compared to a randomly generated population. \citet[][hereafter \citetalias{GilFab20}]{GilFab20}, in an attempt to assess system-level trends, have recently suggested to use several descriptive measures in order to characterise the arrangements of planetary masses, periods, and mutual inclinations within exoplanetary systems. 

In the second approach one quantifies the difference between system architectures using some distance metric. \cite{Alibert19} introduced such a metric, based on planet properties, and then used it to infer the similarity of the protoplanetary discs in which the planetary systems had formed. The metric introduced by \citeauthor{Alibert19} involves representing the planets of the two systems as points on the logarithmic radius-period plane, spreading the points with a Gaussian kernel whose weights are determined by the planet masses, and integrating the squared difference between the two functions.

In this work we pursue the second approach and introduce a new metric to quantify the difference between planetary systems, based on the \textit{Energy Distance}. Originally, the energy distance was introduced as a statistical distance between probability distributions defined on vector spaces \citep{Szekely02}. We propose to consider the distribution of mass among the planets in the examined parameter space (e.g. mass and period) as a discrete probability density function and introduce a weighted extension of the energy distance we henceforth abbreviate as WED. 

Extending on our newly introduced method to quantify the difference between two planetary system architectures, we also propose a method to compare between sets and catalogues of planetary systems, based on WED. We call this extension: ICED (Inter-Catalogue Energy Distance). Such a method can be useful in the context of forward models for population synthesis of planetary systems, such as SysSim \citep{Heetal19}, EPOS \citep{Muldersetal18} or NGPPS \citep{Emsenhuberetal20, Schleckeretal20}. Quantifying the similarity of catalogues or simulated sets of planetary systems is usually done \citep[e.g.][]{Muldersetal18,Heetal19,Emsenhuberetal20,Heetal20} by the Kolmogorov-Smirnov test statistic \citep[KS;][]{Kolmogorov33,Smirnov48} and Anderson-Darling test statistic \citep[AD;][]{AndersonDarling52} to compare between the distributions of planet properties, and with a Pearson $\chi^2$ test or Cressie-Read power divergence \citep[CRPD;][]{CressieRead84} to compare the distribution of planet multiplicity number \citep{Muldersetal18, Heetal19}. We propose in this work to use ICED as a tool to quantify the similarity between two catalogues of planetary systems. 

In Section~\ref{sec:WED} we introduce the detailed computation of the Weighted Energy Distance (WED) metric and present a few examples. In Section~\ref{sec:Kepler}, we calculated a WED matrix for a sample of \textit{Kepler} multiplanetary systems and examine it in the context of the descriptive quantities suggested by \citetalias{GilFab20}. Next, we apply machine-learning exploratory tools on the \textit{Kepler} dataset with the WED metric, looking for suggestive trends in the data.
In section~\ref{sec:ICED} we use the WED to ICED as a unique summary statistic in comparing between catalogues of multiplanet systems. 
Finally, in Section~\ref{sec:Discussion}, we discuss our approach and its applicability.

\section{Weighted energy distance} \label{sec:WED}

The concept of statistical distance is a central theme in statistics. Statistical distances quantify the difference between two statistical objects, which can be random variables, sample distributions, population distributions, etc. A statistical distance may or may not obey all the metric properties, i.e.\ positive definiteness, symmetry and the triangle inequality. Metrics induce certain topological properties on the space on which they are defined, and thus it is usually prefereed that statistical distances would be proper metrics. However, the well-known Kullback-Leibler divergence \citep{KullbackLeibler51}, also known as relative entropy, is a counter example of a statistical distance which is not a metric, since it violates the requirement of symmetry.

In 2002 the statistician G\'{a}bor Sz\'{e}kely introduced a novel statistical distance, which he dubbed "energy distance" \citep{Szekely02}. Energy distance does obey the metric axioms, and it can be applied as a statistical distance between samples or populations defined on any metric space, e.g. high-dimensional Euclidean spaces. Since its introduction, the energy distance served as the basis for a new class of methods to quantify various statistical concepts ranging from goodness of fit to cluster analysis \citep{SzekelyRizzo17}.

Let $f$ and $g$ be the probability density functions of the two independent random variables $X$ and $Y$, defined on the $d$-dimensional Euclidean space (${\mathbb R}^d$). The squared energy distance between the distributions $f$ and $g$ is defined by:

 \begin{equation}
    D^2(f,g) = 2E\Vert X-Y\Vert_d - E\Vert X-X'\Vert_d - E\Vert Y-Y'\Vert_d
	\label{eq:ED}
\end{equation}
where $E$ denotes expected value, and a primed random variable $X'$ denotes an
independent and identically distributed (i.i.d.) copy of $X$, i.e. $X$ and $X'$ are i.i.d., and similarly, $Y$ and $Y'$ are i.i.d. \citep{SzekelyRizzo17}. 

The description above is somewhat abstract, since it relates to the \textit{population} properties of the random variables $X$ and $Y$. Estimating the energy distance from \textit{sample} data is straightforward. Let the data consist of two sets of samples in a certain Euclidean space: $X=\{x_i\}_1^n$ and $Y=\{y_i\}_1^m$. The energy distance between the two distributions from which the two sets are drawn is estimated by:
\begin{equation}
\begin{split}
D^2(X,Y) = \frac{2}{n m} \sum_{i=1}^n \sum_{j=1}^m \Vert x_i-y_j \Vert - \frac{1}{n^2}\sum_{i=1}^n \sum_{j=1}^n \Vert x_i-x_j \Vert - \\
\frac{1}{m^2}\sum_{i=1}^m \sum_{j=1}^m \Vert y_i-y_j \Vert
\end{split}
\end{equation}
The first term in the expression above is actually proportional to the average of all pairwise distances between samples in $X$ and samples in $Y$. Besides reflecting the distance between the two samples, this average is also undesirably inflated by the variances of the two sample, and the subtraction of the second and third terms removes this unwanted contribution. As \citeauthor{SzekelyRizzo17} show, this definition amounts to a proper metric under very broad conditions.

\citeauthor{SzekelyRizzo17} also show that for one-dimensional random variables, the energy distance between distributions is closely related to the Cram\'{e}r-von~Mises-Kolmogorv distance \citep[e.g.][]{Dar1957} (in Astronomy another very similar distance is commonly applied, which is the Kolmogorov-Smirnov distance). However, the Cram\'{e}r type distances (including Kolmogorov-Smirnov distance) all rely on using the cumulative probability function, which in general Euclidean spaces becomes difficult to define and requires various arbitrary assumptions. The energy distance, on the other hand is straightforward to calculate and is easily applicable to any kind of metric space.

Having established that the energy distance is an easily applicable distance between distributions, we can now, based on the above definition, quantify the distance between two planetary systems $X$ and $Y$ consisting of $n$ and $m$ planets respectively. As most multiplanetary systems currently known have been detected by the \textit{Kepler} mission and the transit method, their two most robustly observable properties are the orbital period $P$ and planet radius $R_{\mathrm{p}}$.
We therefore represent every planetary system as a set of points on the logarithmic period-radius plane, such that each point corresponds to a planet. We also assign to each point a weight corresponding to the planet mass $M_{\mathrm{p}}$, using a mass-radius relation \citep[e.g.][]{Bashietal17}. The need for a weight that would represent the planetary mass, as was suggested also by \citet{Alibert19}, can be demonstrated by a hypothetical case of two systems of giant planets that are otherwise identical except for the existence of an additional very small planet in one of them. The weight function guarantees that the two systems would still be considered close by WED.

Based on the above we can now define the energy distance between two planetary architectures as follows:
\begin{equation}
    D^2_{\mathrm{ps}}(X,Y) = 2A-B-C \ ,
	\label{eq:WED}
\end{equation}
where in our case A, B, and C are weighted averages of pairwise distances:
\begin{equation}
    A =  \frac{1}{\sum_{i=1}^{n}w_i^X  \sum_{j=1}^{m}w_j^Y}  \sum_{i=1}^{n}\sum_{j=1}^{m}  w_i^X  w_j^Y \Vert x_i-y_j\Vert
	\label{eq:A}
\end{equation}
\begin{equation}
    B = \frac{1}{(\sum_{i=1}^{n}w_i^X)^2} \sum_{i=1}^{n}\sum_{j=1}^{n}  w_i^X  w_j^X \Vert x_i-x_j\Vert
	\label{eq:B}
\end{equation}
\begin{equation}
    C = \frac{1}{(\sum_{i=1}^{m}w_i^Y)^2} \sum_{i=1}^{m}\sum_{j=1}^{m}  w_i^Y  w_j^Y \Vert y_i-y_j\Vert
	\label{eq:C}
\end{equation}
We calculate the WED ($D_{\mathrm{ps}}$) between planetary systems on the $\log P$--$\log R_{\mathrm{p}}$ plane where we use planetary masses (approximated by a simple M-R relation) as the weights $w_i$ in order to assign higher weight to more massive planets.

\begin{figure}
\centering
	\includegraphics[width=10cm, height=3.5cm]{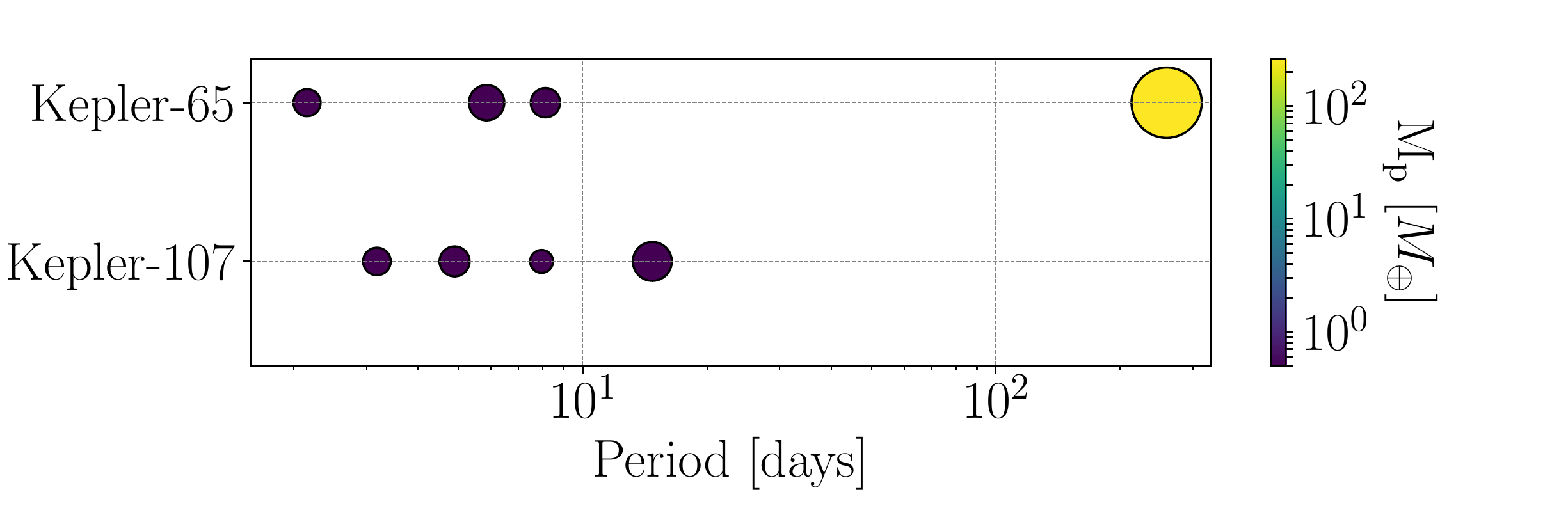}
    \caption{Planet properties of the systems Kepler-107 and Kepler-65, which serve as the starting points of the tests presented in the text. Circle size represents the planet radius in logarithmic scale, while color represents planet mass. In all cases, mass estimates were based on an empirical mass-radius relation \citep{Bashietal17}, except for the planet Kepler-65e (yellow large point) were we used the mass estimates of \cite{Millsetal19}.}
    \label{fig:WEDSys}
\end{figure}

\begin{figure*}
	\includegraphics[width=\columnwidth]{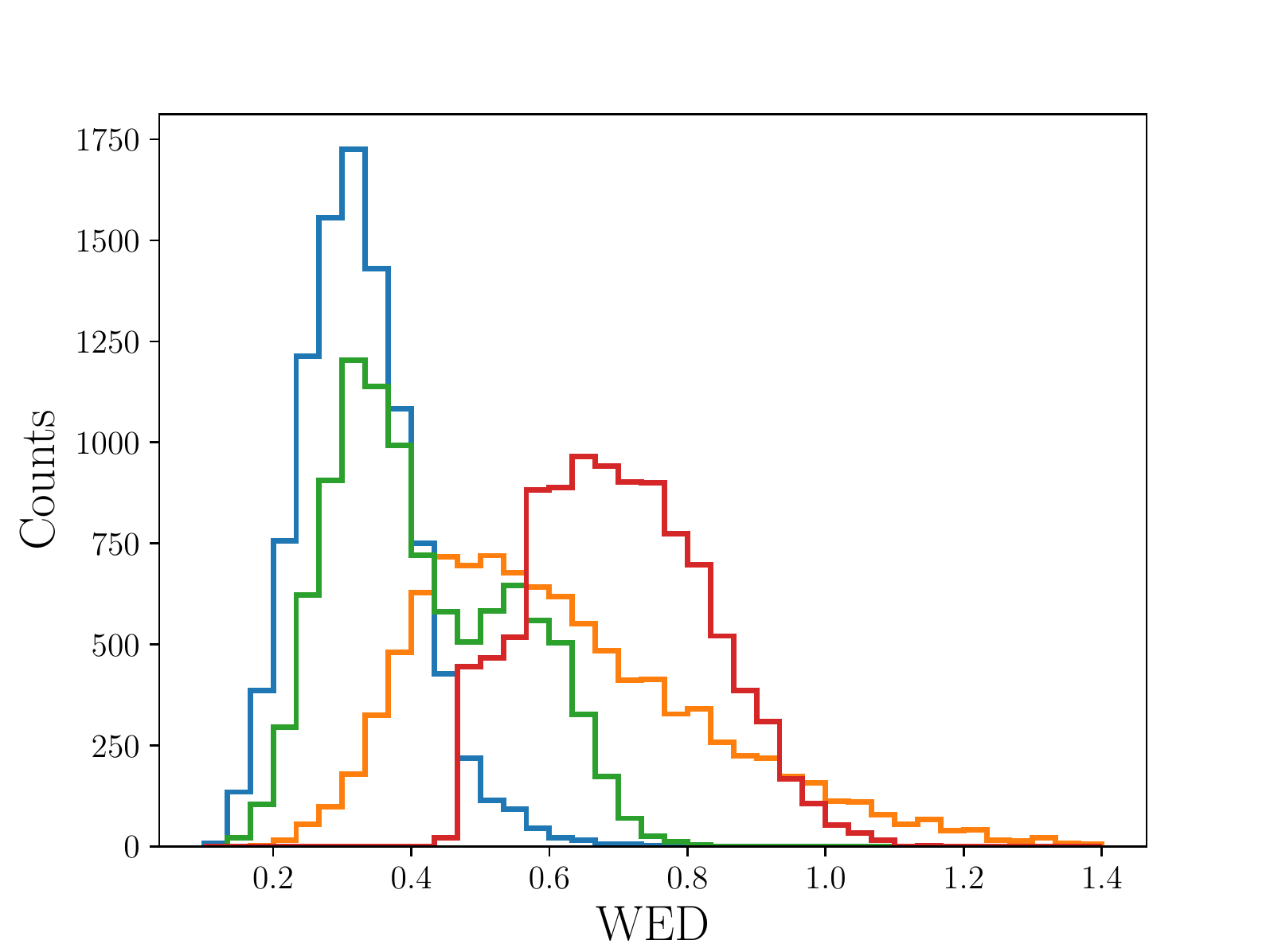}
	\includegraphics[width=\columnwidth]{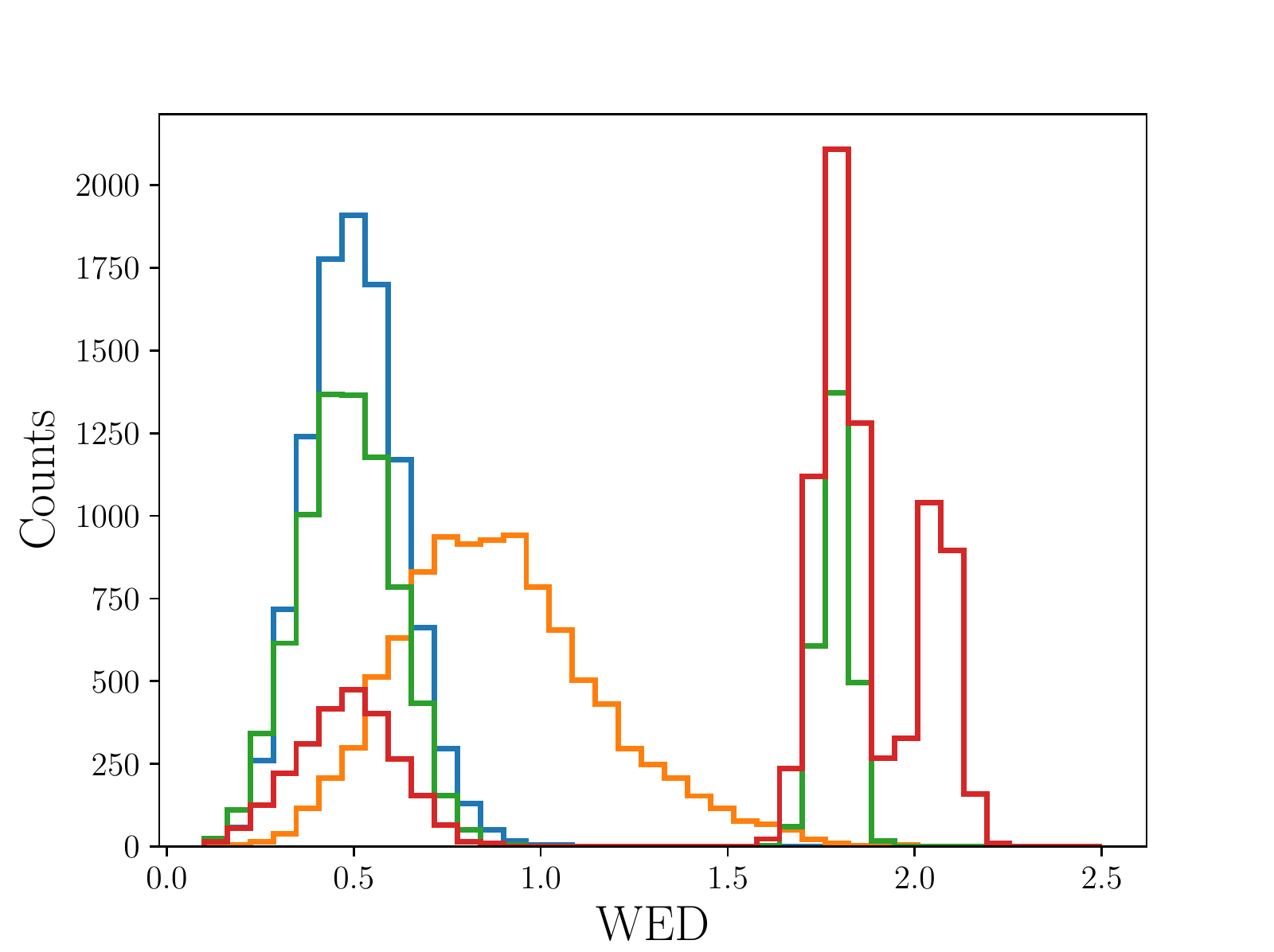}
    \caption{Histograms of WED values for the four tests described in the main text. Each histogram represents $10000$ simulated architectures that deviate in some way from the original four-planet system of Kepler-107 (left panel) and Kepler-65 (right panel). Blue and orange: The properties of all four planets are drawn from a log-normal distribution around the original values, with a standard deviation of $0.1$ and $0.3$, respectively. Green: same as the blue histograms, but with only three of the four original planets, where the excluded planet is chosen randomly. Red: only one of the original planets is retained, chosen at random, with its properties again drawn from a log-normal distribution with a standard deviation of $0.1$.}
    \label{fig:WEDTest}
\end{figure*}

We show several tests we have performed in order to demonstrate the capability of WED to quantify the difference between planetary systems. We have selected the two four-planet systems Kepler-107 \citep{Roweetal14} and Kepler-65 \citep{Chaplinetal13} as the starting points of our tests (Fig. \ref{fig:WEDSys}). We chose Kepler-107 as an example of a tightly-packed system, with four planets of similar radii and masses (except for Kepler-107d which is smaller). Kepler-65 represents another type of multiplanetary system, which includes a distant giant planet, Kepler-65e \citep{Millsetal19}. In each test we calculated the WED between a certain planetary system and each one of a set of $10000$ simulated systems that differed from the original system in a certain random but controlled way. The resulting sets of WED values are shown as histograms in Fig.~\ref{fig:WEDTest}, and can be used as a sanity check to show that the values behave as expected.

In the first test we compiled the $10000$ random architectures by randomly varying the $\log P$ and $\log R_{\mathrm p}$ of all four planets using a Gaussian distribution with a standard deviation $\sigma=0.1$. The WED values of this test are shown in the figure as the blue histogram. In the second test we followed the same procedure, but increased the standard deviation we used in the random simulation to $\sigma=0.3$. The results of this second test are presented as the orange histogram. As expected, in both Kepler-107 and Kepler-65 tests, the WED values are typically larger and also exhibit a wider spread than those in the first test. 

In the third test we changed the architecture into a three-planet architecture, by excluding one randomly-chosen planet and randomly varying the properties of the remaining three planets in the same way as in the first test. The result is shown as the green histogram. In the Kepler-107 system, interestingly, this histogram seems to be bimodal. This is expected when we recall that one of the planets (Kepler-107d) is significantly smaller than the other three, and excluding it leaves a system which is not very different from a similar four-planet system. This explains the left peak of the histogram, which somewhat overlaps the blue histogram. The right peak is caused by excluding one of the three more 'influential' planets. Similarly, in the Kepler-65 test, the clear bimodality is related to the inclusion or exclusion of the dominant giant planet Kepler-65e in the random generation of each simulated system.

In the fourth test the difference from the original system was already drastic -- we changed the system into a single-planet system by randomly drawing this planet out of the four original ones and varying its properties in the same way as in the first and third test. The set of resulting WED values is represented by the red histogram. The WED values are now typically much larger than those of the other tests, again, as expected.

\section{Kepler multiplanet systems} \label{sec:Kepler}

Now that we have formalised WED as a method to quantify the difference between architectures of planetary systems, we can apply WED to the set of \textit{Kepler} multiplanet systems. The sample we use here is the sample of all high-multiplicity ($N_\mathrm{p} \geq 3$) \textit{Kepler} systems \citepalias{GilFab20}, which is based on the California Kepler Survey (CKS) catalogue \citep{Johnsonetal17, Weissetal18}, and comprises $N=129$ systems. We therefore started by calculating the $129 \times 129$ matrix of WED distances of system pairs:
\begin{equation}
a_{i,j} = D_{\mathrm{ps}}(X_i,X_j)
	\label{eq:dcor_a}
\end{equation}

\subsection{Comparison with complexity measures} \label{sec:Dependence}

As mentioned above, \citetalias{GilFab20} introduced a set of descriptive measures, which they dubbed 'complexity measures', to describe the details of planetary system architectures. We set to test whether the information encapsulated by those measures is also contained in the WED distance matrix. We examined the following six measures suggested by \citetalias{GilFab20}\footnote{See section $3$ of \citetalias{GilFab20} for further details.}. We do not include in our analysis the flatness measure of \citetalias{GilFab20}, since our current simple definition of WED does not include the inclination.:

\begin{enumerate}
	\item \textit{Dynamical mass} -- $\mu$ -- describes the overall mass scale of the system;
	\item \textit{Mass partitioning} -- $Q$ -- captures the variance in masses;
	\item \textit{Monotonicity} -- $M$ -- quantifies the size ordering of the planets;
	\item \textit{Characteristic spacing} -- $S$ -- the average separation between planets in units of mutual Hill radii;
	\item \textit{Gap complexity} -- $C$ -- summarises the relationships between orbital periods of adjacent planets;
	\item \textit{Multiplicity} -- $N_{\mathrm{p}}$ -- is the number of observed planets in a
    system.
\end{enumerate}

\citetalias{GilFab20} have already tested the dependence among the various complexity measures using the distance correlation dependence metric \citep[dCor;][]{Szekelyetal07, SzekelyRizzo17}. Unlike the Pearson correlation coefficient, which tests for linear relationships between two variables, dCor can be used to quantify nonlinear associations (dependence) between variables in arbitrary metric spaces. dCor vanishes if and only if the random variables are statistically independent, and it assumes the value $1$ only for strict linear relation (either positive or negative). The usefulness of dCor has already been demonstrated in some recent astrophysical works (\cite{Martetal14, Zucker18}; \citetalias{GilFab20}). 

In our case, the metric space we focus on is the space of planetary system architectures, endowed with the metric defined by WED. Estimating dCor requires first calculating a distance matrix for the two variables in question, which we have already calculated for the $129$ high-multiplicity systems $a_{i,j}$. The distance matrix for each of the six complexity measures is simply calculated using the simple absolute difference:
\begin{equation}
b_{i,j} = |y_i-y_j|
	\label{eq:dcor_b}
\end{equation}
The next step in the dCor calculation is ‘double centring’ of the matrices $a$ and $b$:
\begin{equation}
A_{i,j} = a_{i,j} - a_{i,.} - a_{.,j} + a_{.,.} ; 
B_{i,j} = b_{i,j} - b_{i,.} - b_{.,j} + b_{.,.}
	\label{eq:dcor_centring}
\end{equation}
where $a_{i,.}$ is the \textit{i}-th row mean, $a_{.,j}$ is the \textit{j}-th column mean, and $a_{.,.}$
is the grand mean of the matrix $a$. Similar definitions apply for the matrix $b$.

\begin{figure*}[htb]
\centering
  \begin{subfigure}[b]{.345\linewidth}
    \centering
    \includegraphics[width=.99\textwidth]{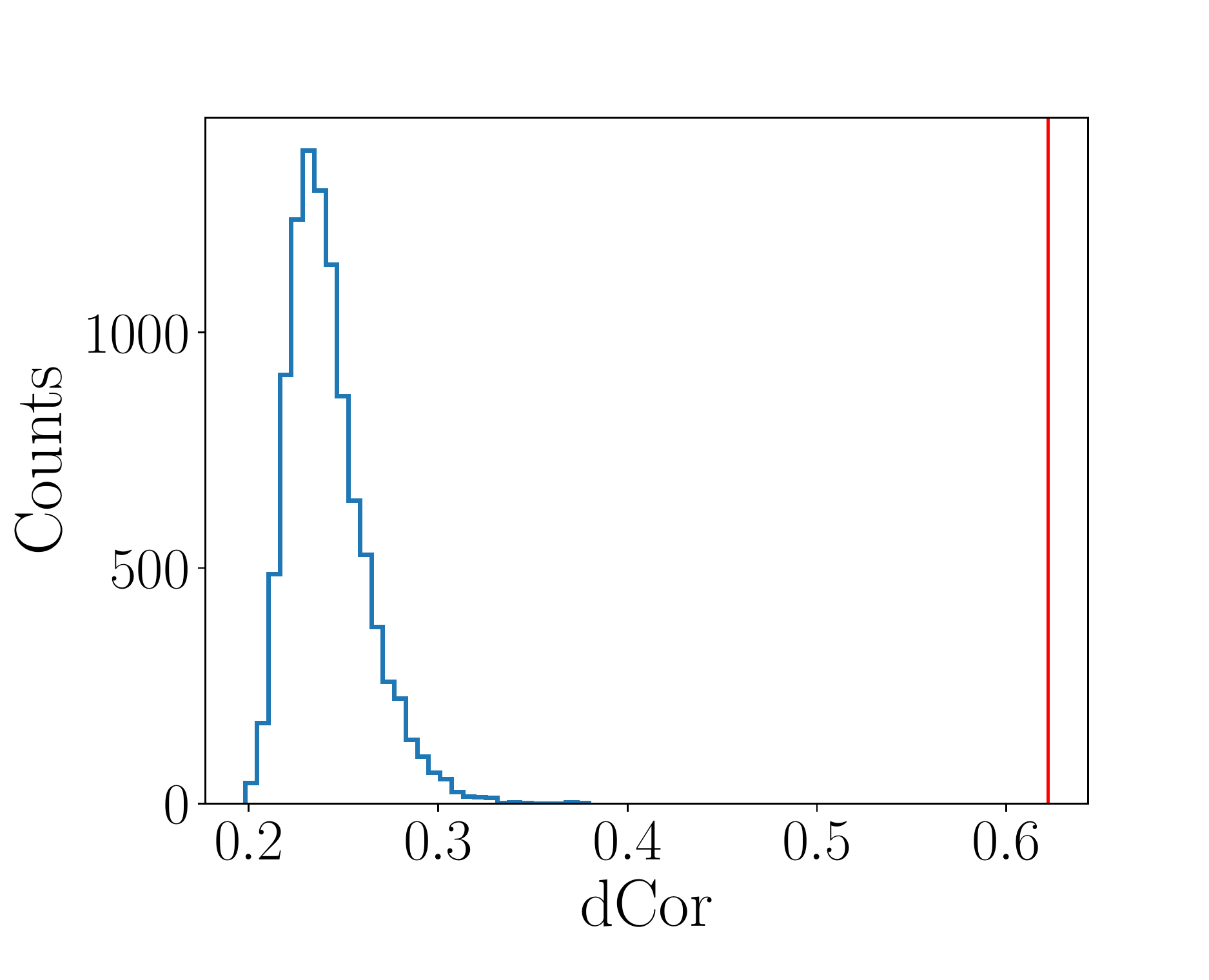}
    \caption{Dynamical mass ($\mu$)}\label{fig:1a}
  \end{subfigure}%
    \begin{subfigure}[b]{.345\linewidth}
    \centering
    \includegraphics[width=.99\textwidth]{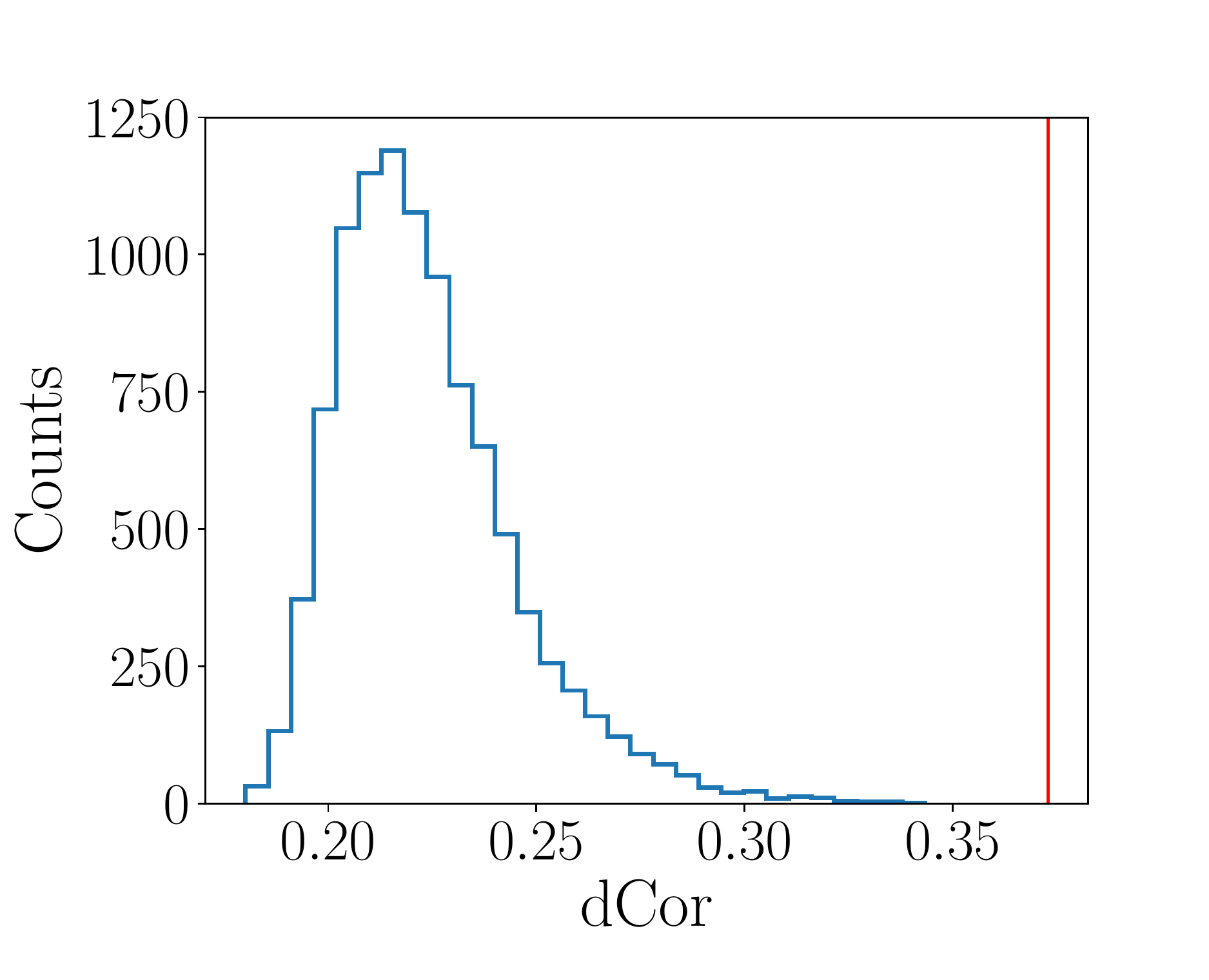}
    \caption{Mass partitioning ($Q$)}\label{fig:1a}
  \end{subfigure}%
  \begin{subfigure}[b]{.345\linewidth}
    \centering
    \includegraphics[width=.99\textwidth]{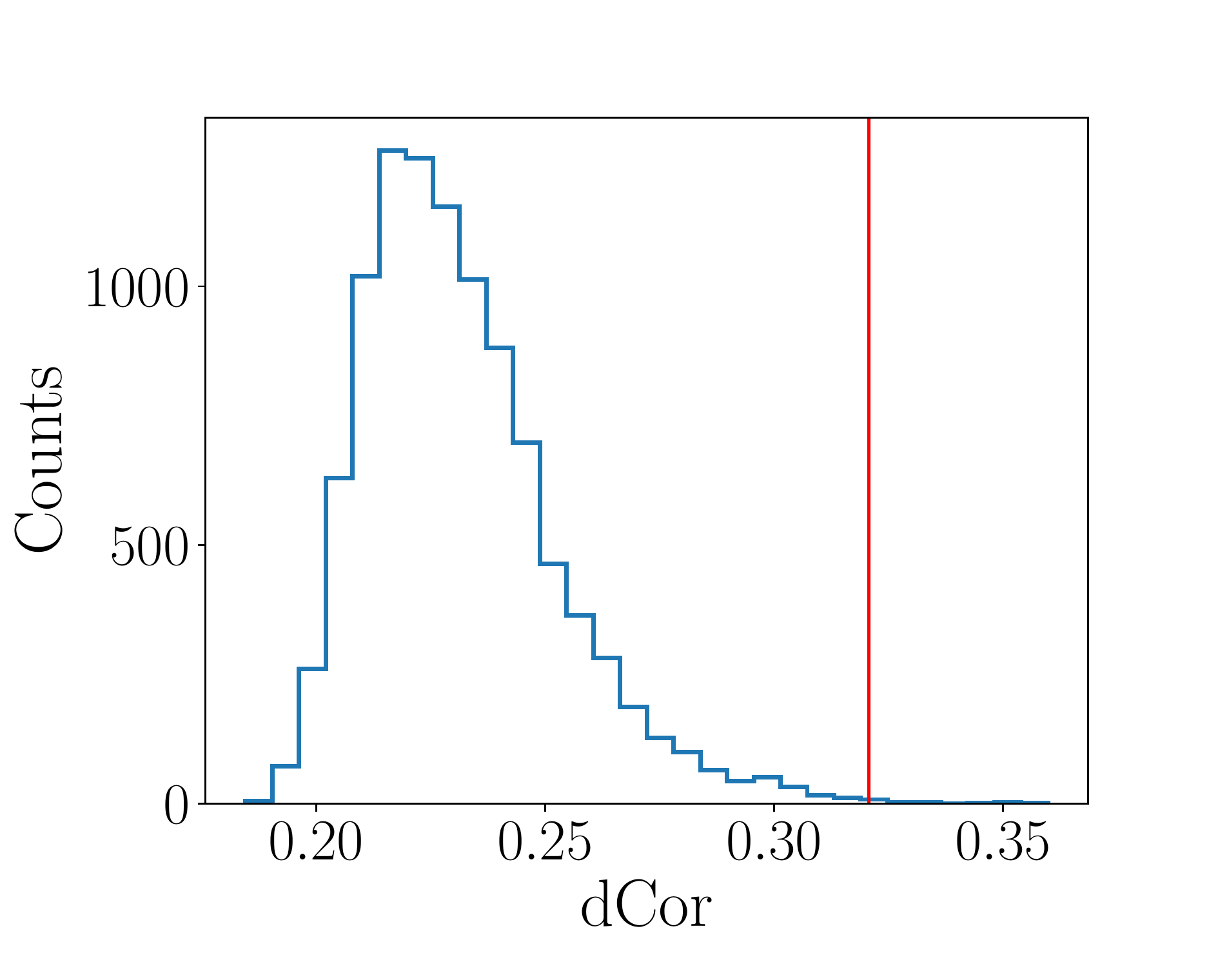}
    \caption{Monotonicity ($M$)}\label{fig:1b}
  \end{subfigure}\\%
  \begin{subfigure}[b]{.345\linewidth}
    \centering
    \includegraphics[width=.99\textwidth]{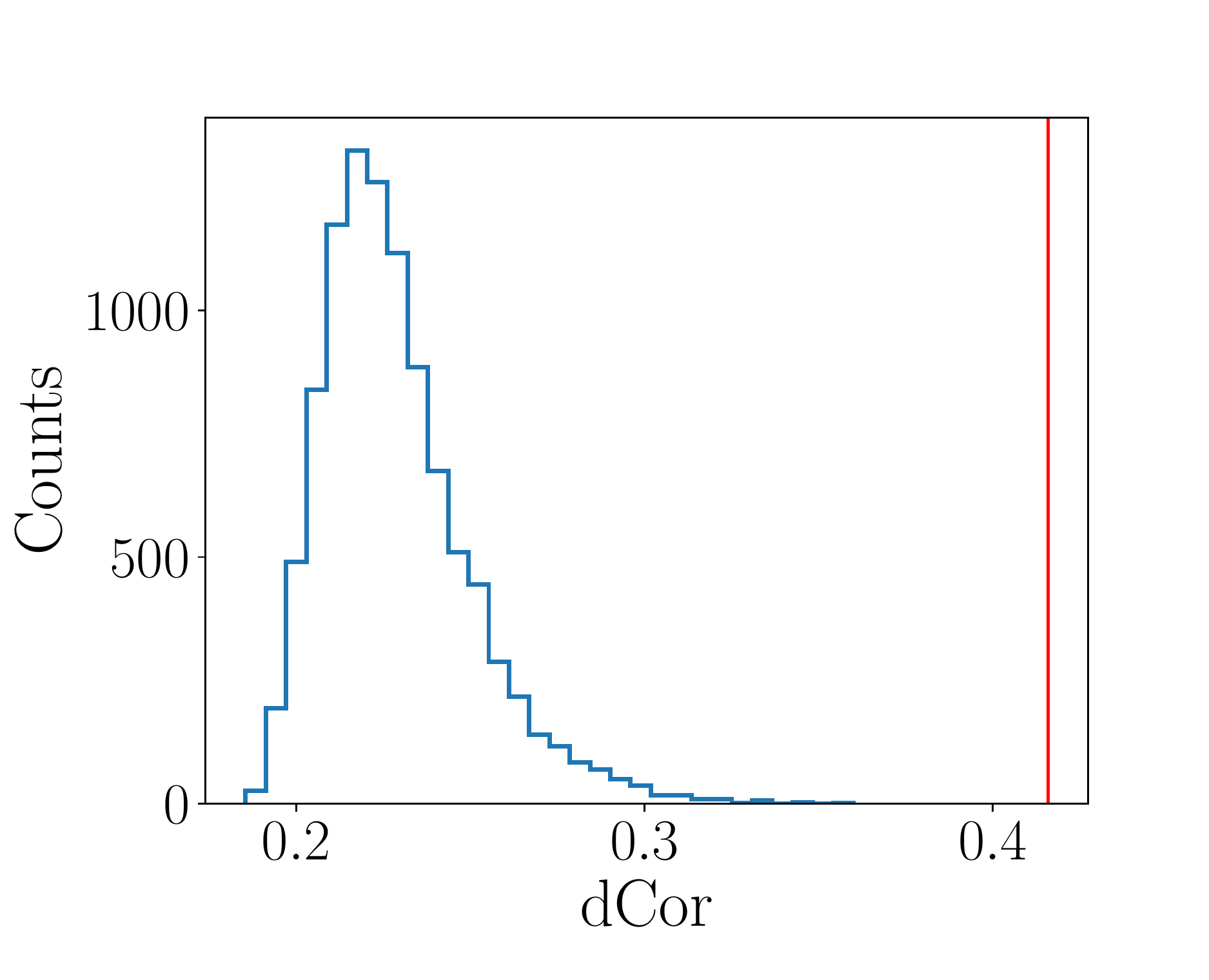}
    \caption{Characteristic spacing ($S$)}\label{fig:1a}
  \end{subfigure}%
  \begin{subfigure}[b]{.345\linewidth}
    \centering
    \includegraphics[width=.99\textwidth]{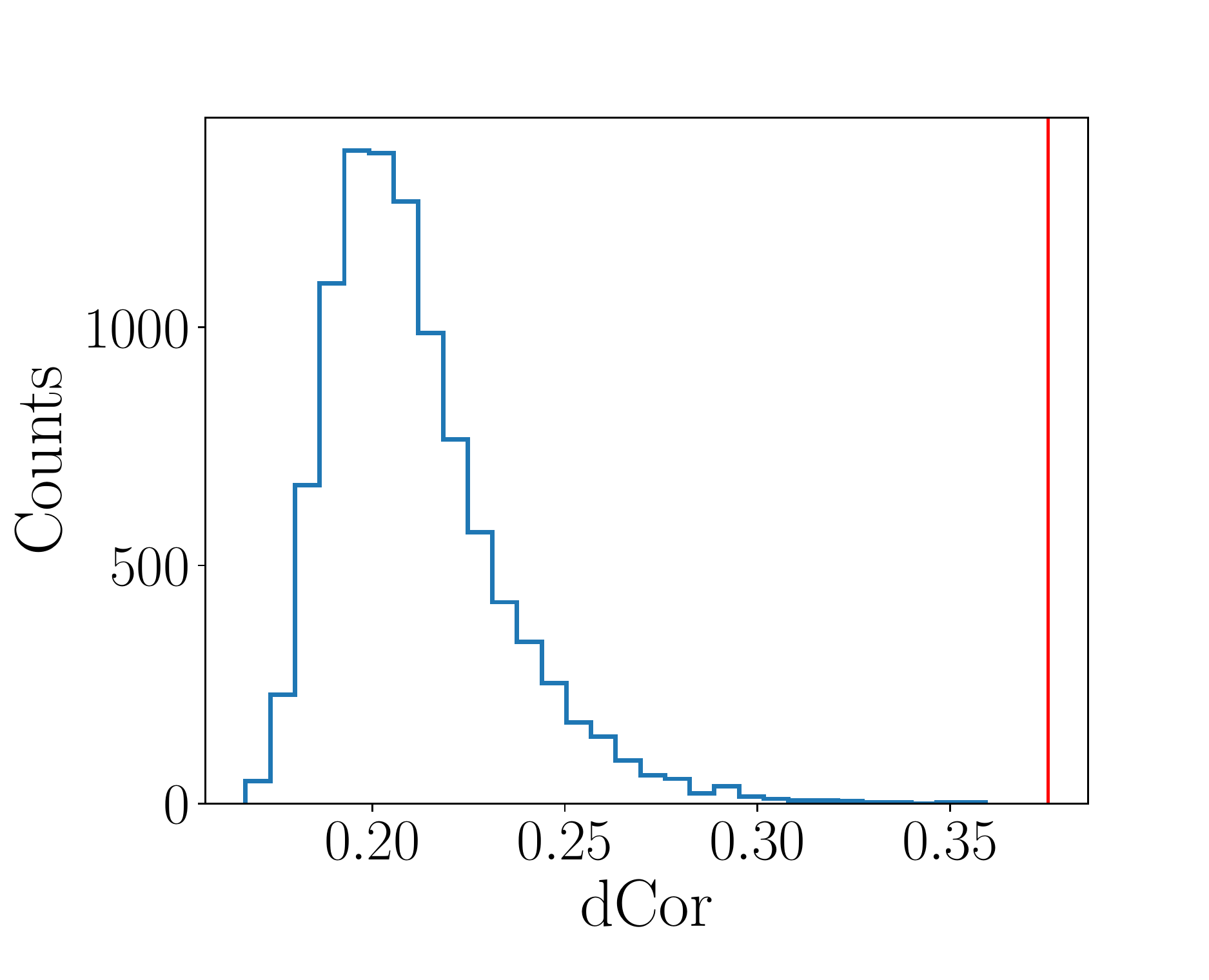}
    \caption{Gap complexity ($C$)}\label{fig:1b}
  \end{subfigure}%
   \begin{subfigure}[b]{.345\linewidth}
    \centering
    \includegraphics[width=.99\textwidth]{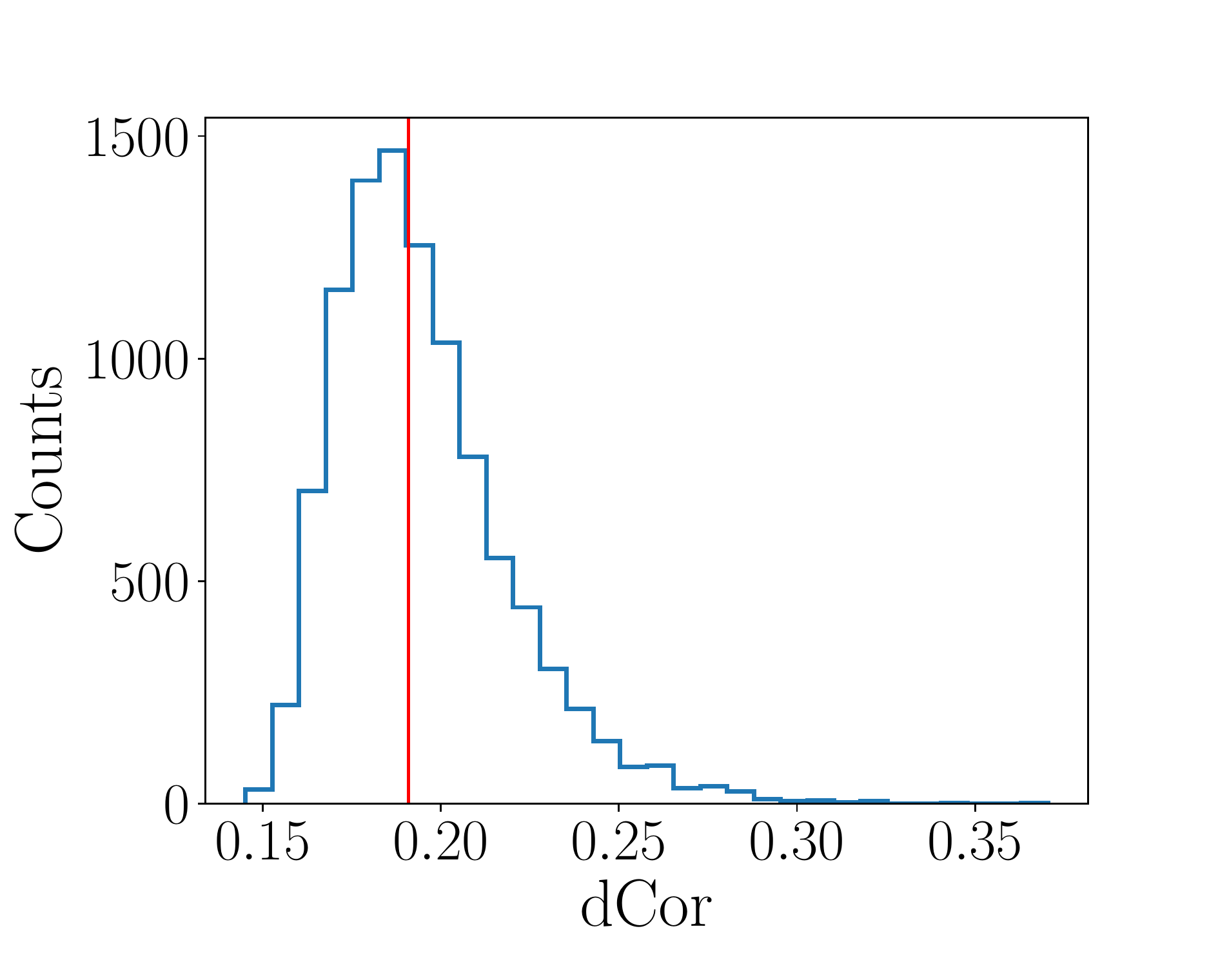}
    \caption{Multiplicity ($N_{\mathrm{p}}$ )}\label{fig:1b}
  \end{subfigure}%
  
  \caption{Distribution of distance correlation values (dCor) between \textit{Kepler} high multiplicity systems with WED and $10000$ random shuffles of the six \citetalias{GilFab20} measures. The vertical red line marks the dCor value of the original ordering.}\label{fig:DC_WED}
\end{figure*}

We thus calculated six dCor values, as measures of dependence between the metric space of architectures with WED and each complexity measure using the expression:
\begin{equation}
\mathrm{dCor}^2 = \frac{\sum_{i,j} A_{i,j} B_{i,j}}{\sqrt{(\sum_{i,j} A_{i,j}^2)(\sum_{i,j} B_{i,j}^2)}}
	\label{eq:dcor}
\end{equation}
In order to assess the significance of these six dCor values, we repeatedly calculated them with $10000$ random shufflings of the corresponding complexity measures. We then counted how many of the dCor values were higher than those of the original ordering. We present in Fig.~\ref{fig:DC_WED} the distribution of the dCor values between the sample of $129$ architectures and the random shuffles of \citetalias{GilFab20} complexity measures. The red line represents the value obtained for the original ordering. 

In all cases, except for the multiplicity, we found that all $10000$ random shufflings resulted in very low dCor values as compared to the original ordering. We therefore can only estimate an upper limit on their p-values of $10^{-4}$ (putting aside the monotonicity p-value of $8 \times 10^{-4}$). Table~\ref{table:dCor_pval} shows the dCor values for all six complexity measures. We therefore conclude that the WED metric effectively contains the combined information encoded by the complexity measures (except for the multiplicity), and maybe even additional information.

\begin{table}
\caption{\label{table:dCor_pval} Distance correlation values and significance for the dependence between architectures with WED and the complexity measures introduced by \citetalias{GilFab20}.}
\centering
\begin{tabular}{lcc}
\hline\hline
  & dCor & p-value\\
\hline
$\mu$ (Dynamical mass)        & $0.62$ & $ <1 \times 10^{-4}$\\
$Q$ (Mass partitioning)      & $0.38$ & $ <1 \times 10^{-4}$\\
$M$ (Monotonicity)           & $0.32$ & $8  \times 10^{-4}$\\
$S$ (Characteristic spacing) & $0.42$ & $ <1 \times 10^{-4}$\\
$C$ (Gap complexity)         & $0.37$ & $ <1 \times 10^{-4}$\\
$N_{\mathrm{p}}$ (Multiplicity)         & $0.19$ & $0.49$\\
\hline \hline
\end{tabular}
\end{table}

\subsection{Exploratory tools} \label{sec:Exploratory}

We now set to further explore the WED matrix of the high-multiplicity \textit{Kepler} systems using exploratory machine-learning tools.

\begin{figure*}
 \centering
\includegraphics[width=9cm]{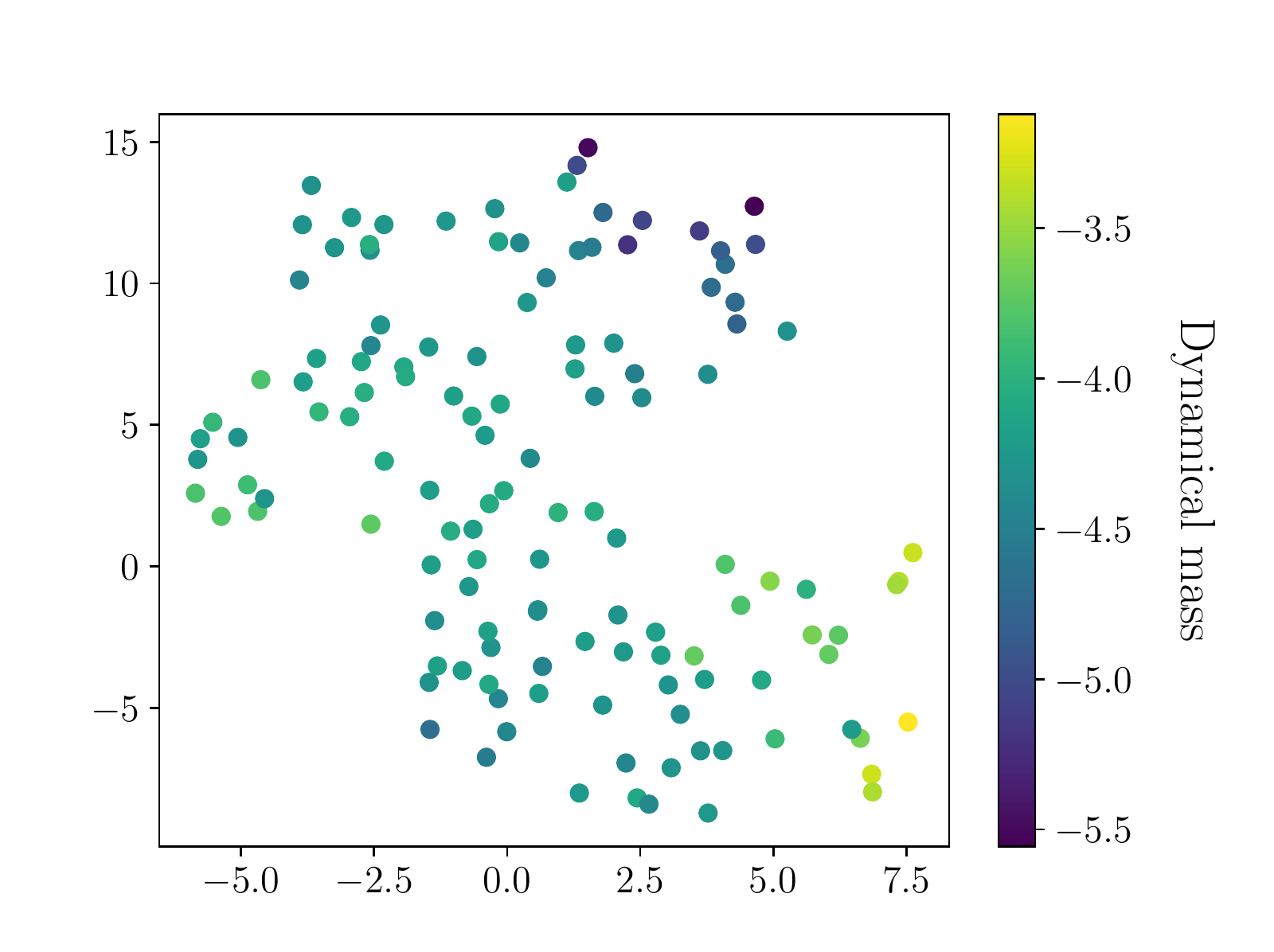}
\includegraphics[width=9cm]{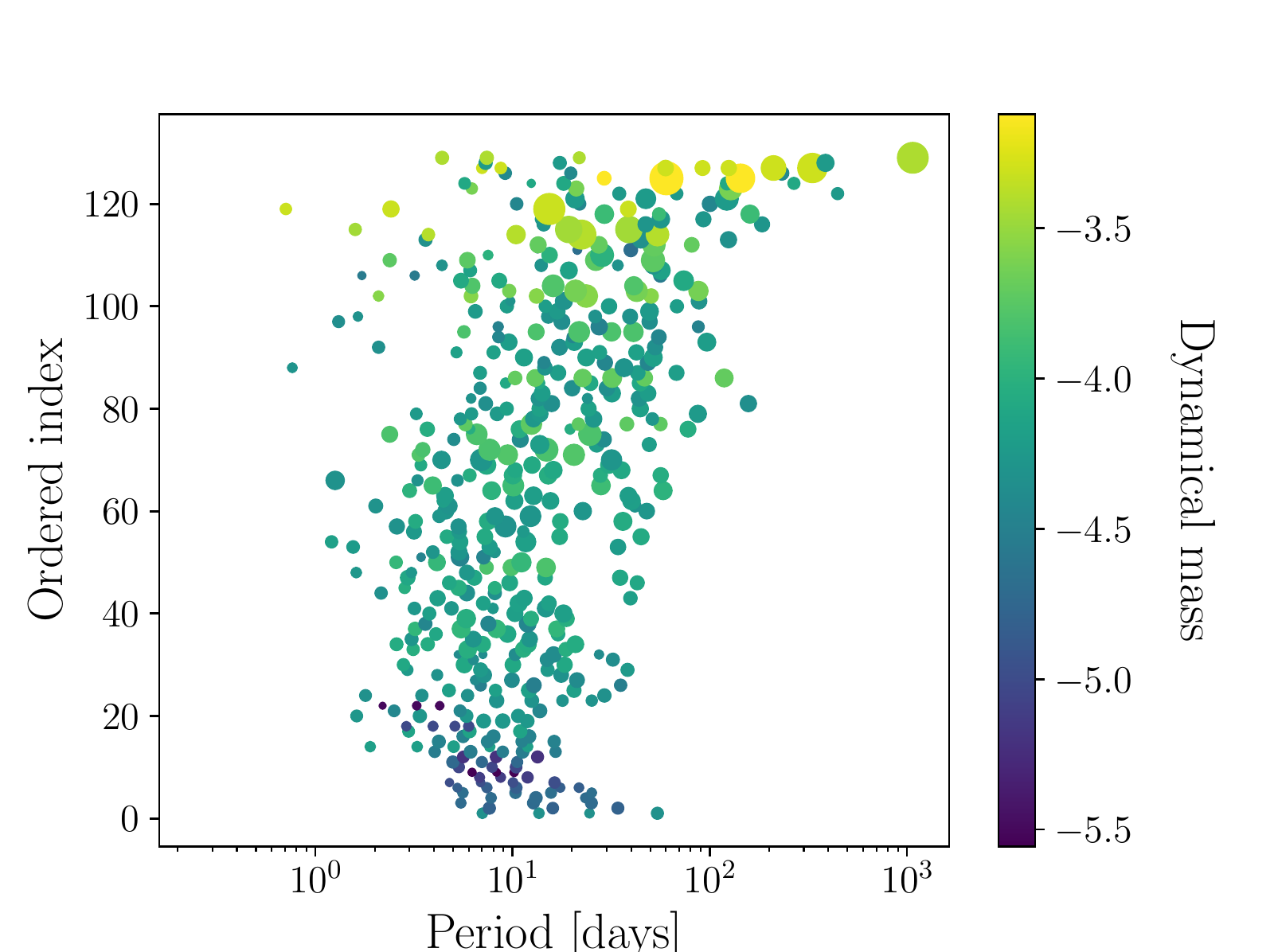}
    \caption{The results of using WED in the context of machine-learning exploratory tools: t-SNE (left) and the Sequencer (right). In both cases, it seems there is a clear trend with the order of systems based on dynamical mas ($\mu$). It is important to note that each point on the t-SNE plot represent a single planetary system, while on the Sequencer plot, each \textit{row} of points represent a single planetary system.}
    \label{fig:Explor}
\end{figure*}

We begin by applying the t-Distributed Stochastic Neighbor Embedding \citep[t-SNE;][]{VanHinton08} method. t-SNE is a tool to visualise high-dimensional data. It converts similarities between data points to joint probabilities and then minimises the Kullback-Leibler divergence \citep{KullbackLeibler51} between the joint probabilities of the low-dimensional embedding and the high-dimensional data (actually, in our case, the data is not high-dimensional bat rather data in a metric space with no clearly defined dimensionality). \cite{Alibert19} applied a similar approach in order to relate the similarity between planetary systems with the similarity of protoplanetary discs in which they formed. In our case we use t-SNE to search for patterns in the set of \textit{Kepler} high-multiplicity systems by trying to identify possible clusters or trends with other properties using colour coding. On the left panel of Fig.~\ref{fig:Explor}, we present the result of performing t-SNE using the default hyperparameters of the scikit-learn python library \citep{Pedregosaetal19} on our sample, colour coded based on the dynamical mass ($\mu$)\footnote{It is important to note that individual axes in t-SNE are not easily interpretable in terms of the original feature space or any physical quantities. Thus one should not try to interpret t-SNE quantitatively but take it mainly as a visualization tool.}. A clear trend is evident, as expected from our dependence tests. Furthermore, the plot may suggest the emergence of a pattern of two clusters. \citetalias{GilFab20} have already suggested the emergence of such two patterns in their attempt to apply path-based spectral clustering \citep[R-PBSC;][]{ChaYeu2008}.

Next, we employ an exploratory tool called the Sequencer, recently introduced by \citet{BaronMenard20}. The Sequencer is an algorithm that searches for trends in a dataset. As a demonstration, \citeauthor{BaronMenard20} showed how the Sequencer can identify the stellar main sequence in a set of stellar spectra based only on the pairwise differences among the spectra and without using any stellar parameter like the temperature. In order to perform its task, the algorithm uses the shape of graphs describing the multi-scale similarities (e.g. in our case using the WED matrix) in the data. In particular, it uses the fact that continuous trends (sequences) in a dataset lead to more elongated graphs. We applied the Sequencer on the same set of $129$ high-multiplicity \textit{Kepler} systems as before, with the WED metric as a measure of dissimilarity, to see whether any order emerges. We applied the algorithm using a Breadth-First Search (BFS) walk, i.e.\ along the longest branch of the graph and scanning each branch along the way. We present the result on the right panel of Fig.~\ref{fig:Explor}. Each circle represents a planet, and its radius is proportional to the planet radius. All planets in the same system share a row, which is also colour coded based on the dynamical mass ($\mu$). It seems that the main obvious feature in the preferred sequence is a progression in terms of the dynamical mass, but also in terms of the typical period. This may also be related to the fact that the dynamical mass and the characteristic spacing are the quantities that show the strongest distance correlation in Table \ref{table:dCor_pval}.

\section{Distance between catalogues} \label{sec:ICED}

WED, as a measure of similarity between planetary systems, can now be used as the basis for a method to compare catalogues or samples of planetary systems, which we henceforth denote Inter-Catalogue Energy Distance -- ICED.

Suppose that $\widetilde{X}$ and $\widetilde{Y}$ are two independent sets consisting of $N$ and $M$ planetary systems respectively. In the same way we have defined WED based on weighted Euclidean distance, we can now define $D^2_{\mathrm{pc}}(\widetilde{X},\widetilde{Y})$, based on the WED between architectures of planetary systems in $\widetilde{X}$ and $\widetilde{Y}$:
\begin{equation}
    D^2_{\mathrm{pc}}(\widetilde{X},\widetilde{Y}) = 2\widetilde{A}-\widetilde{B}-\widetilde{C}
	\label{eq:ICED}
\end{equation}
where $\widetilde{A}$, $\widetilde{B}$, and $\widetilde{C}$ are simply the pairwise
WEDs:
\begin{equation}
    \widetilde{A} =  \frac{1}{NM}  \sum_{i=1}^{N}\sum_{j=1}^{M} D_{\mathrm{ps}}(X_i,Y_j)
	\label{eq:AA}
\end{equation}

\begin{equation}
    \widetilde{B} = \frac{1}{N^2} \sum_{i=1}^{N}\sum_{j=1}^{N} D_{\mathrm{ps}}(X_i,X_j)
	\label{eq:BB}
\end{equation}

\begin{equation}
    \widetilde{C} = \frac{1}{M^2} \sum_{i=1}^{M}\sum_{j=1}^{M} D_{\mathrm{ps}}(Y_i,Y_j)
	\label{eq:CC}
\end{equation}

\begin{figure}
	\includegraphics[width=\columnwidth]{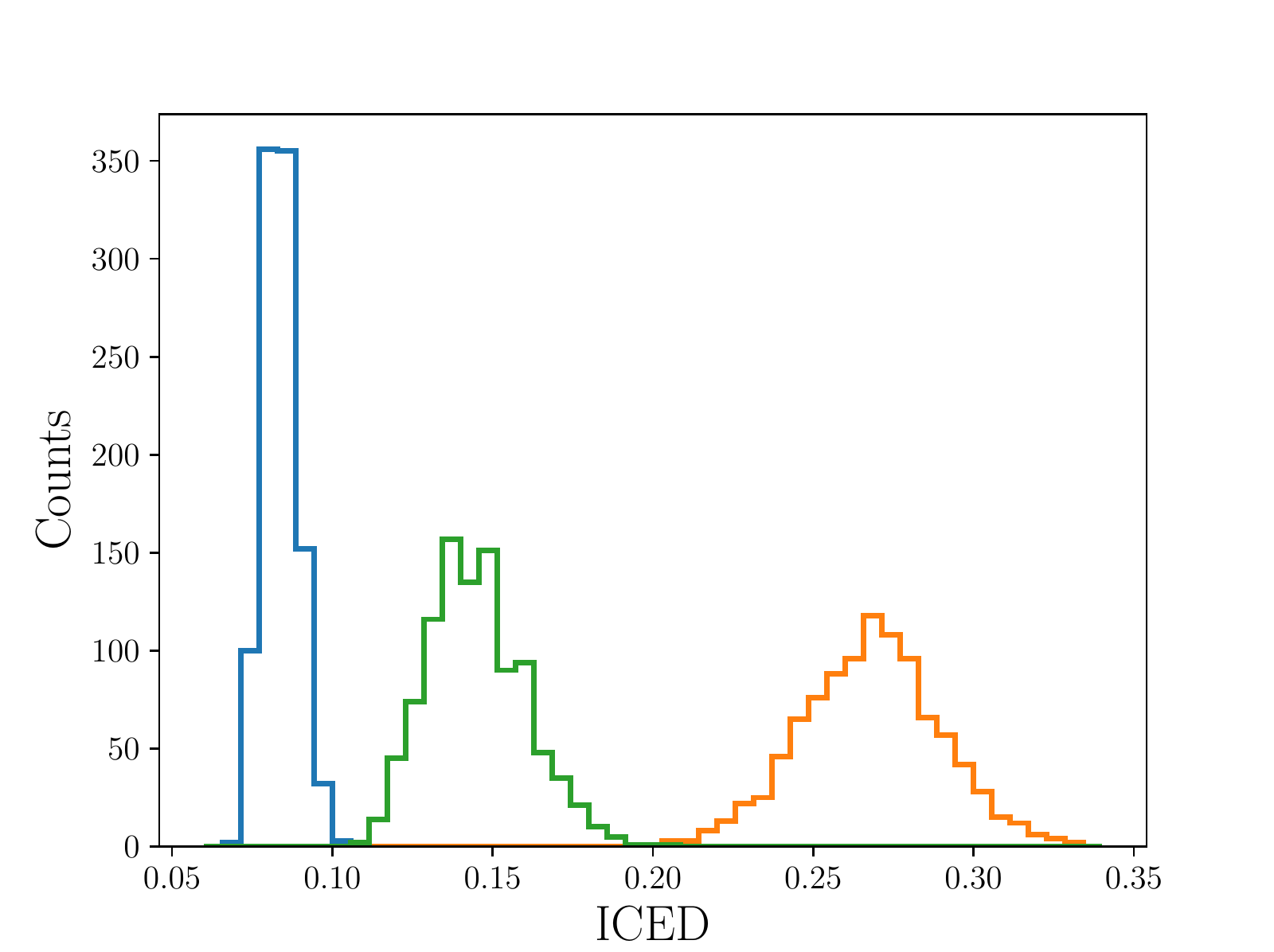}
    \caption{Histograms of ICED values for the three tests described in the main text. Each histogram represents $1000$ simulated catalogues that deviate in some way from the original set of $129$ high-multiplicity ($N_\mathrm{p} \geq 3$) \textit{Kepler} systems. Blue and orange: The properties of all planets are drawn from a log-normal distribution around the original values, with a standard deviation of $0.1$ and $0.3$, respectively. Green: same as the blue histograms but including possibility to change each systems planet multiplicity by drawing the number of planets per system from a uniform distribution between $0$ planets to the actual number of planets detected around that system $N_\mathrm{p}^{i}$.}
    \label{fig:ICEDTest}
\end{figure}

We demonstrate the ability of ICED to quantify the difference between sets of planetary systems by a test similar to the one we used in Sec.~\ref{sec:WED} for WED. We use as a reference sample the sample of $129$ high-multiplicity ($N_\mathrm{p} \geq 3$) \textit{Kepler} systems of Sec. \ref{sec:Kepler}. In each test we calculated the ICED between our reference sample and each one of a set of $1000$ simulated catalogues, which differed from the original one in a certain random but controlled way. The results of these tests are shown as histograms in Fig. \ref{fig:ICEDTest}.

In the first test, we compiled the $1000$ random catalogues by randomly varying $\log P$ and $\log R_{\mathrm p}$ of all planets in each system, using a Gaussian distribution with a standard deviation $\sigma=0.1$. The ICED values of this test are shown in the figure as the blue histogram. In the second test we followed the same procedure, but increased the standard deviation we used in the random simulation to $\sigma=0.3$. The results of this second test are presented as the orange histogram. As expected, the ICED values are typically larger and exhibit a wider spread than those in the first test. Both tests preserved the multiplicity of the systems included in the reference sample.

Next, we tested how the ICED values differed as we changed the systems multiplicities. In the third test (green histogram) we randomly changed the multiplicity by drawing the number of planets in each system from a uniform distribution between $0$ planets (effectively excluding this system as it did not host planets anymore) and the actual number of planets detected in the system. We then randomly chose the remaining planets and varied their properties in the same way as in the first test. We can clearly see that in this case where we allowed the multiplicity to attain lower values (i.e.\ higher discrepancy from the original sample), the ICED values again tended to be higher compared to the first test. However, it seems that ICED in this case is generally lower than that obtained in the case in which we preserved the multiplicity but significantly varied ($\sigma=0.3$) the planet properties (second test). Thus we conclude that WED, as a tool to quantify similarity of two sets of multiplanetary systems, is affected more by the properties of the planets and less by the number of planets in each system (i.e.\ multiplicity).

\section{Discussion} \label{sec:Discussion}

We have introduced the Weighted Energy Distance (WED) as a novel metric based on energy distance to quantify the difference between architectures of planetary systems \footnote{We provide a public Python package (PASSta: Planetary Architecture System Statistics) with implementation of the tools presented in this work at: \url{https://github.com/dolevbas/PASSta}.}. We showed that the information conveyed by WED includes the information conveyed by previous descriptive measures used to characterise the arrangements of planetary systems \citepalias{GilFab20}. Nonetheless, it is important to note that while WED is an exploratory tool that can be applied to discover interesting and previously unknown trends between architectures of planetary systems, the complexity measures of \citetalias{GilFab20} are still very valuable as they provide a clear and explicit description of the system properties, which are conveniently and intuitively interpretable. 

Our aim was to suggest a way to quantify dependencies between the architectures and other characteristics of the planetary system without restricting the analysis to any specific feature of the architecture. In that sense dCor with WED is an integrative way to find such relations, which is agnostic of any predisposition we might have regarding planetary system architectures. Once we detect such a relation, the next step would be to check whether any of the more intuitive quantities like the \citetalias{GilFab20} measures can explain this finding, or maybe we should dig deeper and come up with new insights.

In this context it is important to address our finding that the multiplicity did not exhibit any significant dependence with WED (as manifested in dCor). At first sight this seems a drawback of WED, but at a second look this may not be such a big surprise. As an example, WED will be small for two systems with similar two giant planets that differ only in the number of the other small planets. This can be justified since the physical formation processes that formed these two systems were probably similar, as both formed similar two dominant planets. Alternatively, two systems with an identical set of terrestrial planets except for the replacement of one of the terrestrial planets with a giant planet will rightfully be considered very different. WED will reflect those trends, showing its value in quantifying the difference between architectures in a physically meaningful way. The fact that small planets in wide orbits might easily evade detection affecting dramatically the value of the multiplicity, also casts doubt regarding the physical significance of the multiplicity measure.


We took the WED concept one step further and introduced ICED -- the Inter-Catalogue Energy Distance -- to quantify the difference between samples of planetary systems. We demonstrated how ICED may be used as a summary statistic to compare between catalogues or samples of planetary systems. 
An easy to estimate distance function between architectures of planetary systems can complement forward modeling as a tool to compare between observed and simulated samples of planetary  systems. Previous works \citep{Heetal19, Muldersetal18} used some linear combinations of statistical distances between the separate distributions of the planetary parameters (KS test, AD statistics, Cressie–Read statistics or a simple Euclidean distance). Much like WED, ICED suggests an integrative approach that can be useful also in comparing observed catalogues with planetary population-synthesis models of planet formation and evolution \citep[e.g.][]{Chambers18, Muldersetal19, Emsenhuberetal20}, comparing sub-populations from the same model as was recently used by \cite{Schleckeretal20} and comparing populations produced by varying the details of a model \citep{Nduguetal19}.

Our choice to characterise planetary systems in a two-dimensional space of period-radius (weighted according to planet mass) is not an exclusive choice. Depending on the scientific question at hand and the relevant properties of the architecture or of the catalogue, more dimensions and properties can be added to the definitions of WED and ICED, such as the orbital eccentricity, mutual inclination or the norm of the angular momentum \citep{LaskarPetit17}. For example, if one is interested in exploring systems hosting habitable planets, a change in weights based on planet effective temperatures, may be better suited to quantify the similarity of such systems.


As mentioned above \cite{Alibert19} also presented a metric that used the period and radius quantities. It is therefore instructive to compare the metric of \citeauthor{Alibert19} and the WED. \citeauthor{Alibert19} also converted the locations of the planets on the logarithmic period-radius plane to an analogous probability distributions. However, he effectively used a Gaussian kernel estimation of the probability distribution and 'smeared' the delta functions into Gaussians, with widths that were determined somewhat arbitrarily. Next, he calculated the distance between the two distributions by integrating them on the plane. This involved numerical integration, and furthermore, the boundaries of integration had to be values that were considered effectively infinite. Thus, his metric was sensitive to arbitrary choice of the Gaussian widths, involved approximation of the integral, and was somewhat computationally demanding due to the numeric integration. WED, on the other hand, is quite simple to calculate, and it involves no arbitrary choice of parameters, as its definition (Eqs.~\ref{eq:WED}--\ref{eq:C}) involves only the simplest algebraic operations.

The tools presented in this work are not limited to any specific detection method, and can be generalised to a population of planets detected by any other method. Actually, our method is not limited only to planets and is applicable also for comparison between binary and multiple star systems.

In this work we defined the WED between planetary systems by using a simple Euclidean distance to measure the distance between planets. This can also be generalised by using instead the Mahalanobis distance \citep{Mah1936} in order to account for uncertainties in planet properties, assigning different weights to each property and even account for correlated uncertainties.

As we have shown in Sec.~\ref{sec:Dependence}, presenting the set of possible configurations as a metric space enables the use of the distance correlation. We used this in order to show, as a kind of sanity check, the dependence between the architectures and the complexity measures introduced by \citetalias{GilFab20}. This opens the way to explore possible dependence of the planetary system architectures on various stellar properties, like metallicity, kinematics, age etc. Consequently, using dCor with WED (while accounting for selection effects and detection biases) will allow to explore whether the current properties of planetary systems correlate with their host properties or otherwise, whether the planetary system dynamical evolution erased the information about the primordial conditions in the protoplanetary disks where they formed  \citep{Kipping18, Zhu20}. Thus, instead of testing for such a dependence separately for each of \citetalias{GilFab20} complexity measures, using dCor with WED we will potentially get initial hints about complex relations, which might not be easily detectable through the individual explicit complexity measures.

\begin{acknowledgements} We thank the anonymous referee for the constructive report whose comments helped to improve this manuscript. We are grateful to Dalya Baron for helping us run the Sequencer and further commenting on an early version of the manuscript and to G\'{a}bor S\'{e}kely and Russell Lyons for very insightful discussions about the energy distance and the distance correlation. This research was supported by the ISRAEL SCIENCE FOUNDATION (grant No. 848/16). We also acknowledge partial support by the Ministry of Science, Technology and Space, Israel.
\end{acknowledgements}

\end{document}